\documentclass[aps,pra,twocolumn]{revtex4}

\usepackage{bm,bbm,amsmath,bbold,amssymb,amsbsy,graphicx,amsthm,color,amsthm,hyperref,xfrac,txfonts,mathtools}
\usepackage{fixltx2e}

 \newcommand{\tr}[1]{\text{Tr}}
\newcommand{\ket}[1]{|#1\rangle}
\newcommand{\bra}[1]{\langle#1|}
\newcommand{\sprod}[2]{\langle#1|#2\rangle}
\newcommand{\proj}[1]{\ket{#1}\bra{#1}}
\newcommand{\av}[1]{\langle{#1}\rangle}

\begin{document}
	\title{Exploring corrections to the Optomechanical Hamiltonian}
	\author{Kamila Sala}
	\author{Tommaso Tufarelli}
	\email{tommaso.tufarelli@gmail.com}
	\affiliation{Centre for the Mathematics and Theoretical Physics of Quantum Non-Equilibrium Systems,
		School of Mathematical Sciences, The University of Nottingham,
		University Park, Nottingham NG7 2RD, United Kingdom}
	\begin{abstract}
		{\noindent}We compare two approaches to refine the ``linear model" of cavity optomechanics, in order to describe radiation pressure effects that are beyond first order in the coupling constant. We compare corrections derived from (I) a widely used phenomenological Hamiltonian that conserves the photon number and (II) a two-mode truncation of C. K. Law's microscopic model, which we take as the ``true" Hamiltonian of the system. While these approaches agree at first order, the latter model does not conserve the photon number, resulting in challenging computations from second order onwards. Our numerics suggest that the phenomenological Hamiltonian significantly improves the linear model, yet it does not fully capture all second-order corrections arising from the C. K. Law model. We conclude that, even when the mechanical frequency is much lower than the cavity one, photon number conservation must be eventually given up to model cavity optomechanics with high accuracy.
	\end{abstract}
	\maketitle
	\section{Introduction}
	{\noindent}Cavity quantum optomechanics is a rapidly developing research field exploring the interaction of quantised light with macroscopic mirrors, membranes and levitated nano-objects through radiation pressure \cite{bucco}. The field has recently witnessed impressive experimental progress, including demonstrations of near-ground-state cooling of mechanical oscillators and normal-mode splitting due to phonon-photon interaction \cite{optorevs1}. Among its many applications, optomechanics is a promising platform to demonstrate nonclassical behaviour in massive objects \cite{patre,pulso}, and may provide novel avenues for the coherent manipulation of quantum light \cite{squizzo,Nunne2}. Recently it was also suggested that optomechanics may allow for ultra-high precision measurements, possibly probing Planck-scale corrections to the canonical commutation relations \cite{Pikovski,KumarPlenio}.
	
\begin{figure}
		\includegraphics[width=.9\linewidth]{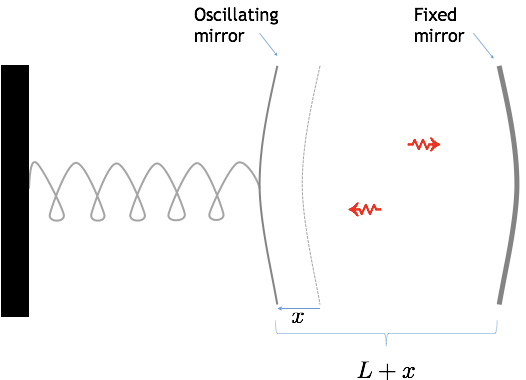}
		\caption{Schematic of an optomechanical setup consisting of a Fabry-Perot resonator with a movable mirror. The length of the cavity, hence its frequency $\omega(x)$, is modulated by the mirror coordinate $x$. The movable mirror is subject to a trapping potential (depicted as a spring), which is usually harmonic to a good approximation. \label{sketch1}}
	\end{figure}

	{\noindent}The simplest optomechanical setup features a single optical cavity mode, whose frequency $\omega(x)$ depends parametrically on the coordinate $x$ (the `position') of a mechanical oscillator \cite{CKLaw}. Perhaps the most notable example is given by a Fabry-Perot resonator with one movable mirror --- see Fig.~\ref{sketch1}.
	
	{\noindent}Most experiments to date have explored the weak coupling regime, in which radiation pressure effects are only visible upon strong driving of the cavity, and the system can be described as a pair of coupled harmonic oscillators \cite{bucco}. Several experimental platforms, however, are now approaching the single-photon strong coupling regime \cite{optorevs1,optorevs2,optorevs3,Nunne1,armour,arxivbig} (strong coupling, or SC, for brevity), in which the anharmonic nature of the radiation pressure interaction must be taken into account. Strong coupling occurs when the the coupling rate of a single photon is greater than the typical loss rates of the setup. Such regime features clear departures from classical behaviour \cite{Qinstab,Nunne1,Nunne2}, and facilitates the production of nonclassical states in both light and mechanics \cite{Bose,ammiraglio}. A further appealing feature of strong coupling optomechanics is that the Hamiltonian can be diagonalized analytically upon linearising the cavity frequency around the origin, as per  $\omega(x)\simeq \omega(0)+x\,\omega'(0)$ (equivalently, this may be seen as a first order expansion in the coupling constant). We shall call \textit{linear model} the resulting Hamiltonian, arguably the most widely used modelling tool in SC optomechanics. 
	
	{\noindent}Despite the undeniable success of the linear model as a theoretical device, the need is arising to go beyond this approximation. For example, it was noted by Brunelli et al. \cite{patre} that the linearized optomechanical Hamiltonian is unbounded from below, with (unphysical) negative energies cropping up at very high photon numbers \cite{thermo}. While this pathology may be seen as somewhat mathematical (see Appendix~\ref{pathology}), it highlights that the model must be eventually refined, particularly in view of future experiments aiming to achieve ever larger coupling strengths. Even in systems that are far from the SC, the quest for ultra-precise measurements (e.g. those pertaining Planck scale physics) \cite{KumarPlenio}, or for the detection of dynamical Casimir effects \cite{Savasto}, demand a more accurate Hamiltonian description of quantum optomechanics.
	
	{\noindent}Armed with these motivations, in this paper we explore and compare two different starting points that may be used to go beyond the linear model: (I) a widely used phenomenological Hamiltonian, which conserves the cavity photon number (\textit{phenomenological approach}); (II) a two-mode truncation of C. K. Law's microscopic Hamiltonian for an optomechanical Fabry-Perot cavity \cite{CKLaw} (\textit{microscopic approach}). We shall take this truncated Law Hamiltonian as the benchmark (or ``true model") against which the various generalizations of the linear model should be judged. In this paper we focus on the realistic situation where the bare cavity frequency is much larger than the mechanical one. We find that approaches I and II agree at first order in the coupling, but at higher orders the microscopic model yields `counter-rotating' terms that violate photon number conservation \cite{Savasto} and make computations challenging. We resort to numerical diagonalization of the Hamiltonian, truncated in Fock space, to deal with such situations (and also to tackle the exact phenomenological Hamiltonian). The numerical examples we explored suggest that the phenomenological approach does a good job in improving the linear model in the typical parameter regimes of optomechanics experiments, yet it does not fully capture the second-order corrections arising from the microscopic treatment. We conclude that photon number conservation must be eventually given up to model cavity optomechanics with high accuracy, even in parameter regimes where dynamical Casimir physics would be usually discounted.

	{\noindent}The paper is organised as follows. In Sec.~\ref{linearSection} we review the linear model of optomechanics. In Sec.~\ref{fenomeno} we show how this model can be refined phenomenologically, while in Sec.~\ref{micro} we discuss model corrections that are microscopically motivated. Sec.~\ref{examples} features numerical examples comparing the various approaches. We draw our conclusions in Sec.~\ref{conclusions}, and in Appendix~\ref{pathology} we discuss in more detail the negative-energies pathology of the linear model which was outlined in \cite{patre}.
	
\section{Linear Model: brief review}\label{linearSection}
{\noindent}We begin by recalling the model that is commonly taken as a starting point in modelling optomechanics. To lighten the notation we take $\hbar=1$, and consider a mechanical mode of unit effective mass. The linear model Hamiltonian reads
\begin{align}
H_{\sf lin}=(\omega_0-G\hat x)\left(\hat a^\dagger \hat a+\frac{1}{2}\right)+\frac{\hat p^2}{2}+\frac12\Omega^2\hat x^2,\label{linearmodel}
\end{align}
where $\hat a$ is the annihilation operator for the cavity field mode, $\hat x$ and $\hat p$ are respectively the canonical position and momentum operators of the movable mirror, $\Omega$ is the bare mechanical frequency, $\omega_0$ is the bare cavity frequency, and the coupling constant $G$ is known as frequency pull parameter. The canonical commutators read  $[\hat a,\hat a^\dagger]=1$, $[\hat x,\hat p]=i,$ while all field operators commute with all mirror operators. In passing we note that the optomechanical interaction strength is often quantified via the vacuum optomechanical coupling strength $g_0=Gx_{\sf zpf},$ where $x_{\sf zpf}=1/\sqrt{2\Omega}$ is the (bare) ground state uncertainty of the mechanical oscillator. Loosely speaking, the dimensionless ratio $g_0/\Omega$ quantifies the influence a single photon can have on the mirror motion. That being said, many of our expressions will have a simpler form when written in terms of $G$. In typical experimental realizations, the hierarchy $\omega_0\gg \Omega\gg g_0$ holds between the model parameters.\\
Hamiltonian \eqref{linearmodel} can be diagonalized exactly \cite{Bose,Nunne1}. The properties of the corresponding eigensystem are well known, and can be obtained from Section~\ref{SECquadratic} by setting $\beta=0$.
\section{Phenomenological Approach}\label{fenomeno}
{\noindent}A first possibility in extending the linear model, often adopted in the literature, is to consider a phenomenological Hamiltonian similar to Eq.~\eqref{linearmodel}, but where the cavity frequency has a generic position dependence $\omega(x)$. Crucially, no change is assumed in the commutation rules for mirror and field operators, an assumption that drastically simplifies computations but may not be justified microscopically \cite{CKLaw} --- we shall elaborate on this point in Section~\ref{micro}.

For the purposes of illustration, from now on we consider an optomechanical setup consisting of a Fabry-Perot resonator with one movable mirror (see Fig.\ref{sketch1}), which amounts to setting $\omega(x)=\omega_0/(1+x/L)$ with $L$ the bare cavity length. However, the reader should keep in mind that the same formalism is applicable when $\omega(x)$ is a generic (positive) function. The phenomenological Hamiltonian is
\begin{align}
H_{\sf phen}=\omega(\hat x)\left(\hat a^\dagger \hat a+\frac{1}{2}\right)+\frac{\hat p^2}{2}+\frac12\Omega^2\hat x^2. \label{Hphenomenal}
\end{align}
If we now pick $L=\omega_0/G=(\omega_0/g_0)x_{\sf zpf}$ and take a first order expansion in $G$ (or equivalently $g_0$), Eq.~\eqref{linearmodel} is recovered. The natural next step is then to retain higher order corrections in $G$. In what follows, we discuss two inequivalent approaches that employ simple harmonic oscillator mathematics to approximate Eq.~\eqref{Hphenomenal} beyond first order. Both methods are in principle amenable to analytical treatment. 
\subsection{Approach 1: photon - controlled harmonic oscillator}\label{refine}
{\noindent}Our first approach involves the recalculation of the mechanical equilibrium position and spring constant as functions of the photon number $n$. Indeed, exploiting the conservation law $[H,\hat a^\dagger \hat a]=0$, we may decompose 
\begin{align}
H_{\sf phen}=\sum_{n=0}^\infty \proj{n}\otimes \left[\frac{\hat p^2}{2}+V_{n}(\hat{x})\right]
\end{align}
featuring the effective $n-$ dependent mechanical potential
	\begin{align}
		V_{n}(\hat{x})&\equiv\omega(\hat x)\left(n+\frac{1}{2}\right)+\frac{1}{2}\Omega^2\hat x^2.\label{potential}
	\end{align}
	The idea is now to perform a quadratic expansion of $V_{n}(\hat{x})$ around its minimum: 
	\begin{align}
		V_{n}(\hat{x})\simeq V_{n}(\bar x_n)+\frac{1}{2}V_{n}''(\bar x_n)\,(\hat x-\bar x_n)^2,
	\end{align}
	where the equilibrium position $\bar x_n$ is implicitly determined by
	\begin{align}
	V_{n}'(\bar x_n)=0.\label{derivata}
	\end{align}
	Note that the above procedure is well defined, since it is easily verified that $V_{n}$ has a unique global minimum at fixed $n$, and that for all $n,$ $V_{n}''(\bar x_n)>0$ (see below). The outlined procedure returns the effective Hamiltonian
	\begin{align}
 H_{\hat n}&\equiv \sum_{n=0}^\infty\proj{n}\otimes \underbrace{\left(\frac{\hat p^2}{2}+\frac{1}{2}\Omega_n^2(\hat x\!-\!\bar x_n)^2+V_{n}(\bar x_n)\right)}_{H_n},\label{Heff}\\
 \Omega_n^2&\equiv V_{n}''(\bar x_n),\label{Omegan}
	\end{align}
	where we have chosen the symbol $H_{\hat n}$ to signify that the photon number operator $\hat n=\hat a ^\dagger \hat a$ controls which harmonic oscillator Hamiltonian $H_n$ is acting on the mirror. For each $n$, the eigenvalues and eigenvectors of $H_{n}$ may be found via straightforward harmonic oscillator techniques. For example the energy eigenvalues, labelled by the cavity photon number $n$ and a mechanical `phonon' number $m$ (both semipositive integers), read 
	\begin{align}
		E_{n,m}=V_n(\bar x_n)+\Omega_n\left(m+\frac{1}{2}\right).\label{eigenenergy}
	\end{align}
	These are all positive, and thus may be used to calculate the partition function without the need for artificial cutoffs \cite{patre}. A fully analytical treatment of $H_{\hat n}$ requires the explicit inversion of Eq.~\eqref{derivata}, which may be recast as
	\begin{align}
	\Omega^2\bar{x}_n=\left(n+\frac12\right)\frac{\omega_0/L}{(1+\bar x_n/L)^2}.\label{crossing}
	\end{align}
{\noindent}This equation for $\bar x_n$ is in principle solvable (it is equivalent to a third-degree polynomial equation) but leads to very cumbersome expressions. A number of observations, however, can be made without resorting to the explicit solution. First, it is evident that it must be $\bar x_n>0$, and a little more thought reveals that $\bar x_n\to\infty$ when $n\to\infty$. Second, notice that Eq.~\eqref{crossing} describes the intersection between a strictly increasing function and a decreasing one, such that for any $n$ there is a unique real solution for $\bar x_n$, as anticipated. Third, implicit differentiation in $n$ shows that $\bar x_n$ is a strictly increasing function of the photon number (as intution about radiation pressure would suggest). Finally, combining Eqs.~\eqref{Omegan} and \eqref{crossing} the photon-dependent mechanical frequencies may be expressed as
\begin{align}
\Omega_n^2=\Omega^2\left[1+\frac{2\bar x_n/L}{(1+\bar x_n/L)}\right],
\end{align}
which is a strictly increasing function of $\bar x_n,$ hence $n$. It is an interestingly mathematical fact that we obtain the elegant limiting value $\lim_{n \to \infty}\Omega_n=\sqrt{3}\Omega$, which does not depend on the cavity parameters $\omega_0$ and $L$. However, this result is unlikely to have any physical significance: the above model may eventually become unreliable when too many photons occupy the cavity (for example, the associated limit $\bar x_n\to\infty$ means that the average cavity length becomes infinite!).
The eigenstates corresponding to $E_{n,m}$ take the form 
\begin{align}
\ket{\Psi_{n,m}}\equiv\ket n\ket{\phi_{n,m}},\label{states}
\end{align}
where the mechanical wavefunctions $\phi_{n,m}(x)=\sprod{x}{\phi_{n,m}}$ are expressible in terms of Hermite functions centered at $x=\bar x_n$ (details not shown). The following eigenstate properties are easily verified:
\begin{align}
&\av{\hat x}_{nm}\equiv\bra{\Psi_{n,m}}\,\hat x\,\ket{\Psi_{n,m}}=\bar x_n,\\
&\Delta x^2_{nm}\equiv\bra{\Psi_{n,m}}(\hat x-\av{\hat x}_{nm})^2\ket{\Psi_{n,m}}=\frac{m+1/2}{2\Omega_n}.
\end{align}	

\begin{figure}[t!]
	\includegraphics[width=\linewidth]{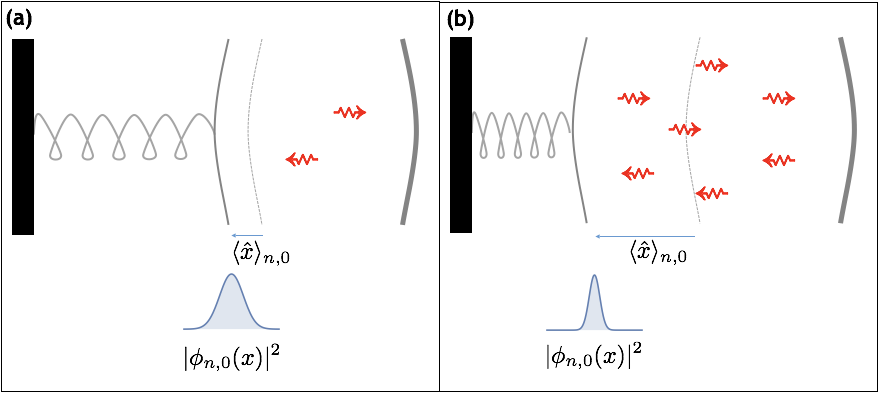}
\caption{{\noindent}In the two analytically solvable models of Section~\ref{fenomeno}, the optomechanical eigenstates $\ket{\Psi_{n,m}}$ possess a qualitatively similar structure. As the photon number is increased going from \textbf{(a)} to \textbf{(b)}, the average mechanical position is pushed further away from the fixed mirror, while the position uncertainty drecreases (i.e., the mechanical wavefunctions get squeezed). Note that the latter squeezing effect is not present in the linear model. \label{squashing}}
\end{figure}

{\noindent}In light of the above discussion we gain the following intuition about the structure of the Hamiltonian eigenstates in Eq.~\eqref{states}: the mechanical wavefunctions $\phi_{n,m}(x)$ are pushed further away from the fixed mirror as the photon number $n$ is increased, while at the same time they become more and more squeezed in position. Such behaviour is schematised in Fig.~\ref{squashing}
\subsection{Approach 2: standard  quadratic expansion}\label{SECquadratic}
{\noindent}Our second approach consists in a more traditional second-order expansion of Eq.~\ref{Hphenomenal}. Specifically, we expand
\begin{align}
\omega(x)\simeq \omega_0-G x+\frac{1}{2}\beta^2 x^2,\label{secondorder}
\end{align}
where $\beta^2=2\omega_0/L^2=2G^2/\omega_0.$ For brevity, we shall refer to the resulting Hamiltonian as the \textit{quadratic model}:
\begin{align}
H_{\sf quad}=\left(\omega_0-G \hat x+\frac{1}{2}\beta^2 \hat x^2\right)\left(\hat a^\dagger \hat a+\frac{1}{2}\right)+\frac{\hat p^2}{2}+\frac12\Omega^2\hat x^2. \label{Hquad}
\end{align}
 While the `quadratic coupling' has been studied before in optomechanics \cite{quadratici}, the focus has often been on the striking differences between a purely linear ($\beta=0$) and a purely quadratic ($G=0$) interaction. Here, on the other hand, our primary goal is to improve the accuracy of our Hamiltonian model: we want to explore the corrections to the linear model due to the higher order $\beta$ term. This is in a similar spirit to Ref.~\cite{KumarPlenio}, where it was argued that the inclusion of higher order terms is necessary to investigate Planck-scale modifications to the canonical commutator. Taking Eq.~\eqref{Hquad} as a starting point, the relevant calculations for the quadratic model follow closely those of Sec.~\ref{refine}, \textit{mutatis mutandis}. What is more, we can here obtain compact analytical expressions. Specifically, we may recast the Hamiltonian in the form \eqref{Heff}, with parameters
	\begin{align}
	\Omega_n^2&=\Omega^2+\left(n+\frac{1}{2}\right)\beta^2,\label{quadOmega}\\
		\bar x_n&=\frac{G(n+1/2)}{\Omega_n^2},\label{Quadxn}
	\end{align}
	so that the resulting energy eigenvalues read
	\begin{align}
	E_{n,m}=\left(n+\frac{1}{2}\right)\omega_0+\Omega_n\left(m+\frac{1}{2}\right)-\frac{G^2(n+1/2)^2}{2\Omega_n^2},\label{energy}
	\end{align}
	valid for all semipositive integers $n,m$. Also here we shall discuss the limit $n\to\infty$ as a matter of mathematical curiosity. Note that the quadratic model predicts a finite average mechanical displacement even in the limit of large $n$, while the mechanical frequencies  are divergent with asymptotic scaling $\Omega_n \propto\sqrt{n}$, in turn implying that the mechanical wavefunctions become infinitely squeezed in position:
	\begin{align}
	&\lim_{n \to \infty}\av{\hat x}_{nm}=\lim_{n \to \infty}\bar x_n= \frac{G}{\beta^2},\\
	&\lim_{n \to \infty}\Omega_n=+\infty,\\
	&\lim_{n \to \infty}\Delta x_{nm}=0.
	\end{align}
	\section{Microscopic approach}\label{micro}
	{\noindent}In this section we take as a starting point the microscopic Hamiltonian derived by C. K. Law through canonical quantization of a Fabry-Perot cavity with a movable mirror \cite{CKLaw}. For the sake of tractability we shall assume that only two degrees of freedom, one optical and one mechanical, are sufficient to model our optomechanical system. Under this assumption, the Law Hamiltonian takes the simplified form
	\begin{align}
	H_{\sf mic}=\frac{\hat \Pi^2}{2}+\frac{1}{2}\omega^2(\hat x)\hat Q^2+\frac{\hat p^2}{2}+\frac12\Omega^2\hat x^2, \label{HCKLaw}
	\end{align}
	where $\hat Q,\hat \Pi$ are canonical quadrature operators for the cavity field mode and $\omega(\hat x)$ retains the same form as before. The elementary canonical commutator for the cavity field is now $[\hat Q,\hat \Pi]=i$, and again all field operators commute with the mechanical variables $\hat x$ and $\hat p$. Note that Hamiltonian \eqref{HCKLaw} implies $\tfrac{d}{dt} {\hat p}\propto \hat Q^2$, i.e., the instantaneous radiation pressure force is now proportional to the intensity of a single field quadrature (rather than to the field energy). This makes sense: since the (transverse) electric field $\mathbf E$ must vanish on the mirror, the radiation pressure force will be proportional to the squared magnetic field $|\mathbf B|^2$ alone (evaluated at the mirror surface).	Hence, within the two-mode approximation, we argue that Eq.~\eqref{HCKLaw} should be a more reliable description of radiation pressure physics than Hamiltonian \eqref{Hphenomenal}. In the next section, Eq.~\eqref{HCKLaw} will play the role of benchmark, or ``true model", against which our various approximations shall be compared. In the present section, we instead discuss the corrections to the linear model that arise in this framework, highlighting the issues that were not present in the phenomenological approach.  
	Note that Hamiltonian \eqref{Hmic} does not yield photon number conservation in general. More precisely, the concept of photon itself relies on identifying a reference frequency, for which a natural choice is $\omega_0\equiv\omega(0)$. We then choose the following definition for the cavity annihilation operator:
	\begin{align}
	\hat a=\sqrt{\frac{\omega_0}{2}}\hat Q+\frac{i}{\sqrt{2\omega_0}}\hat \Pi.\label{aref}
	\end{align}
	{\noindent}Since $\hat a$ commutes with all mirror operators, it allows us to define a Fock space for the cavity field that does not depend on the mirror coordinate $\hat x$. We found this approach convenient to set-up numerical calculations based on the truncation of Fock spaces. In passing we note that this was not the path taken in the seminal paper by C. K. Law \cite{CKLaw}, where instead $\hat x-$ dependent annihilation operators and Fock spaces were introduced to describe the cavity field. Hamiltonian \eqref{HCKLaw} may be recast as
	\begin{align}
	H_{\sf mic}&=\nu(\hat x)\left(\hat a^\dagger \hat a+\frac12\right)+\!\frac{\lambda(\hat x)}2\left(\hat a^2\!+\!\hat a^{\dagger 2}\right)+\frac{\hat p^2}{2}+\frac12\Omega^2\hat x^2\label{Hmic},	\end{align}
	where $\nu(\hat x)=(\omega(\hat x)^2+\omega_0^2)/{2\omega_0},\, \lambda(\hat x)=(\omega(\hat x)^2-\omega_0^2)/{2\omega_0}$ and the cavity photon number $\hat a^\dagger \hat a$ is evidently not conserved due to the appearance of counter-rotating terms (i.e. $\hat a^2+\hat a^{\dagger2}$). Let us now discuss how to approximate Hamiltonian \eqref{HCKLaw} at first and second order in the coupling $G$. A straightforward first-order expansion of Eq.~\eqref{Hmic} yields
\begin{align}
H_{\sf mic}^{(1)}= H_{\sf lin}-\frac{G\hat x}2\left(\hat a^2\!+\!\hat a^{\dagger 2}\right)\simeq H_{\sf lin},\label{microlinear}
\end{align}
where the linear model \eqref{linearmodel} has been recovered through a rotating-wave approximation, justifiable in the parameter regime $\omega_0\gg\Omega\gg g_0$ that we are considering here. More in detail, a back-of-the-envelope calculation shows that the counter-rotating terms only provide a contribution at order ${\cal O}\left({G^2\hat x^2}/{\omega_0}\right)$ --- for example this can be proved by time-averaging the Hamiltonian on timescales $\gg1/\omega_0$ \cite{James}. In the parameter regimes of interest we can thus reach the reassuring conclusion that the microscopic and phenomenological approaches yield the same linear model. We should however point out that there can be situations where $\omega_0$ and $\Omega$ have comparable magnitude, and in that case the $(\hat a^2+\hat a^{\dagger2})$ term cannot be neglected even at first order in $G$  \cite{Savasto}.

{\noindent}It is at second order in $G$ that non-negligible discrepancies arise between the microscopic and phenomenological approaches. Following similar lines of reasoning as above, we find that the second-order microscopic Hamiltonian is
\begin{align}
H_{\sf mic}^{(2)}= H_{\sf quad}+\frac{\beta^2\hat x^2}4\left(\hat a^\dagger\hat a+\frac12\right)-\frac{G\hat x}2\left(\hat a^2\!+\!\hat a^{\dagger 2}\right), \label{microquad}
\end{align}
where we recall $\beta^2=2\omega_0/L^2=2G^2/\omega_0$, and we note two additional terms that were not present in the phenomenological model and cannot be neglected at the considered order. Due to the counter-rotating terms, Hamiltonian \eqref{microquad} presents the same computational challenges as the original model in Eq.~\eqref{HCKLaw} --- albeit its numerical solution appears to be slightly faster in the examples we explored. The relevance of Eq.~\eqref{microquad} to our discussion is more conceptual than practical: it highlights that the phenomenological approach of Section~\ref{fenomeno}, when taken beyond first order in $G$, may begin to miss some of the finer details of optomechanical physics. In particular, the appearance of counter-rotating terms provides a \textit{qualitative} departure from the phenomenological model.\\
{\noindent}Luckily, the examples to follow indicate that the analytical models derived via the phenomenological approach may still provide significant improvements to the linear model in the regime $\omega_0\gg\Omega\gg g_0$.
%%%%%%%%%%	FIGURONA
\begin{figure*}[t!]
	\begin{flushleft}
		\includegraphics[width=.31\linewidth]{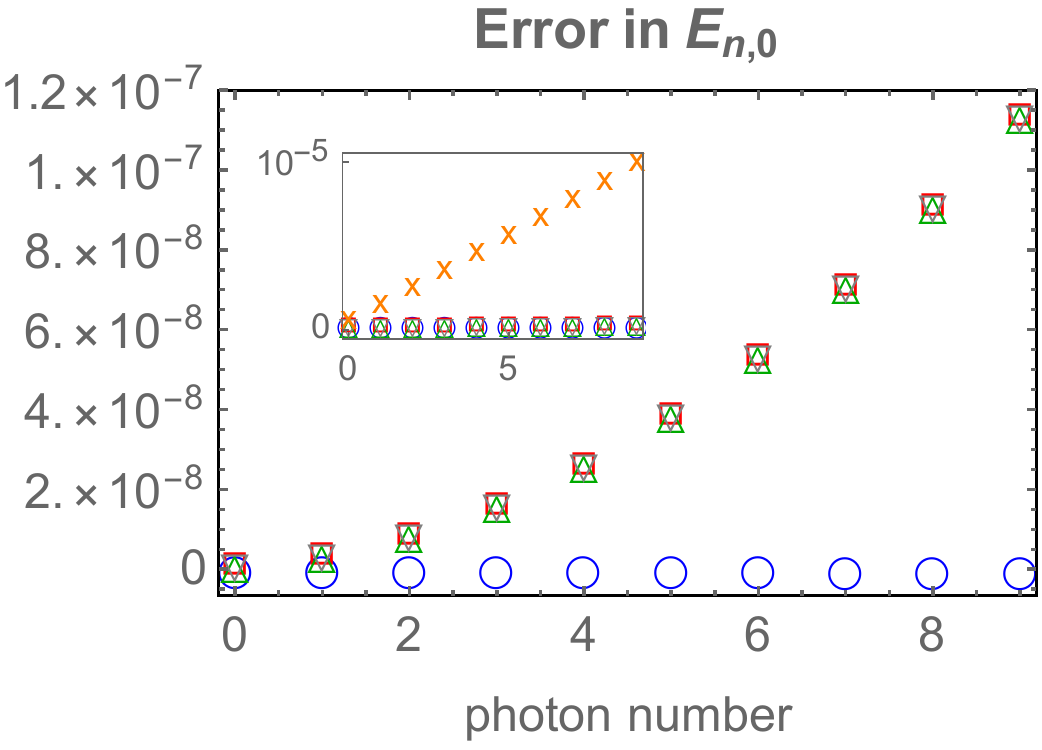}\hspace{.02\linewidth}\includegraphics[width=.31\linewidth]{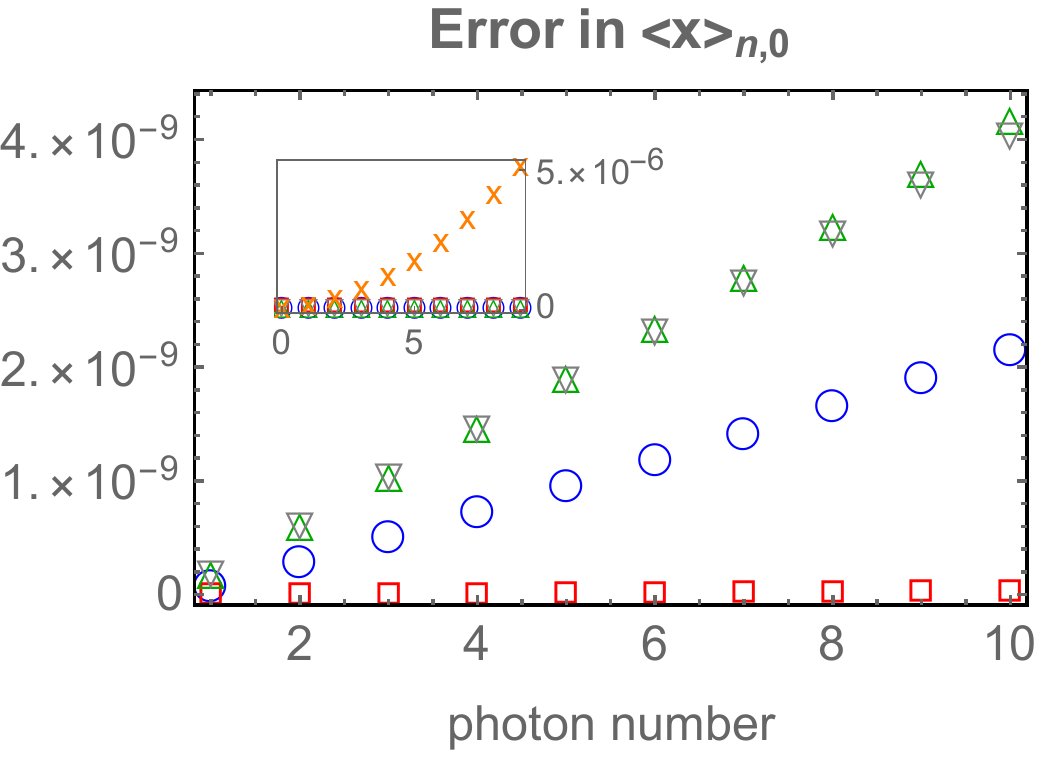}\hspace{.02\linewidth}	\includegraphics[width=.31\linewidth]{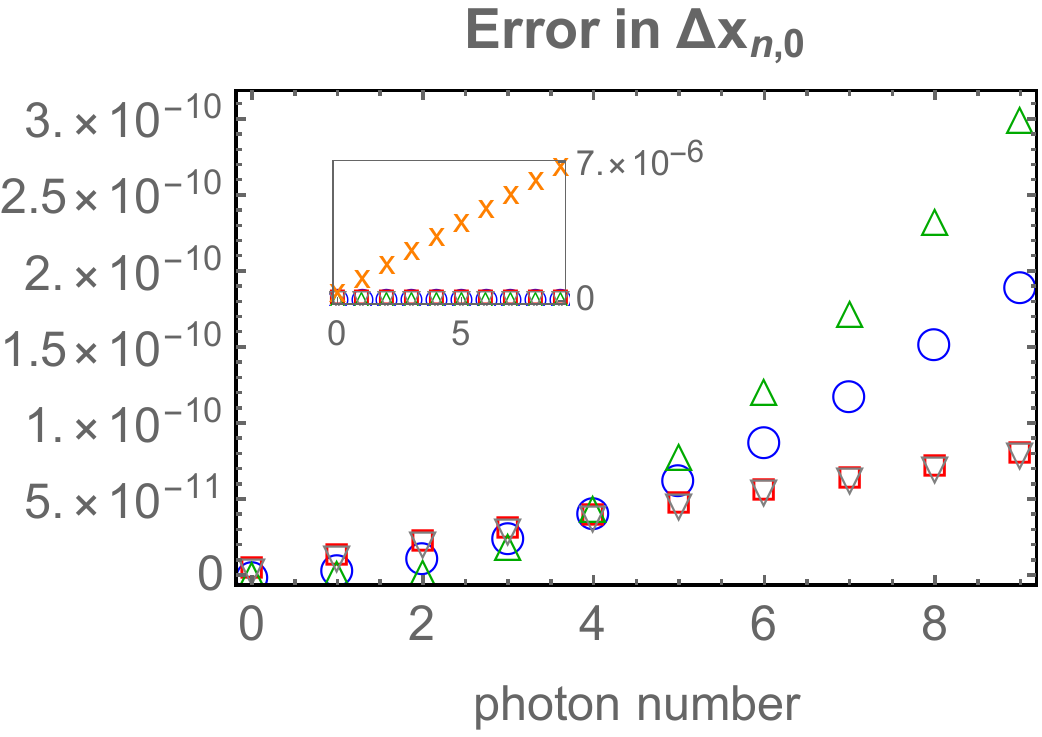}\\
		\hspace{.25cm}\\
		\includegraphics[width=.31\linewidth]{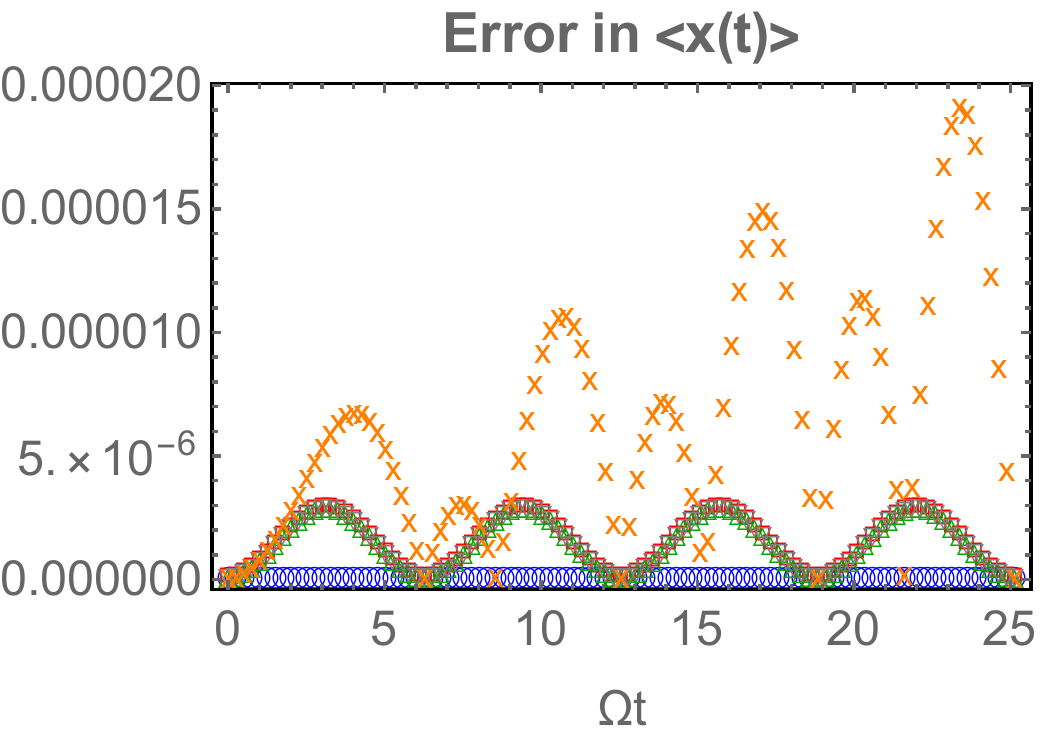}\hspace{.02\linewidth}	\includegraphics[width=.31\linewidth]{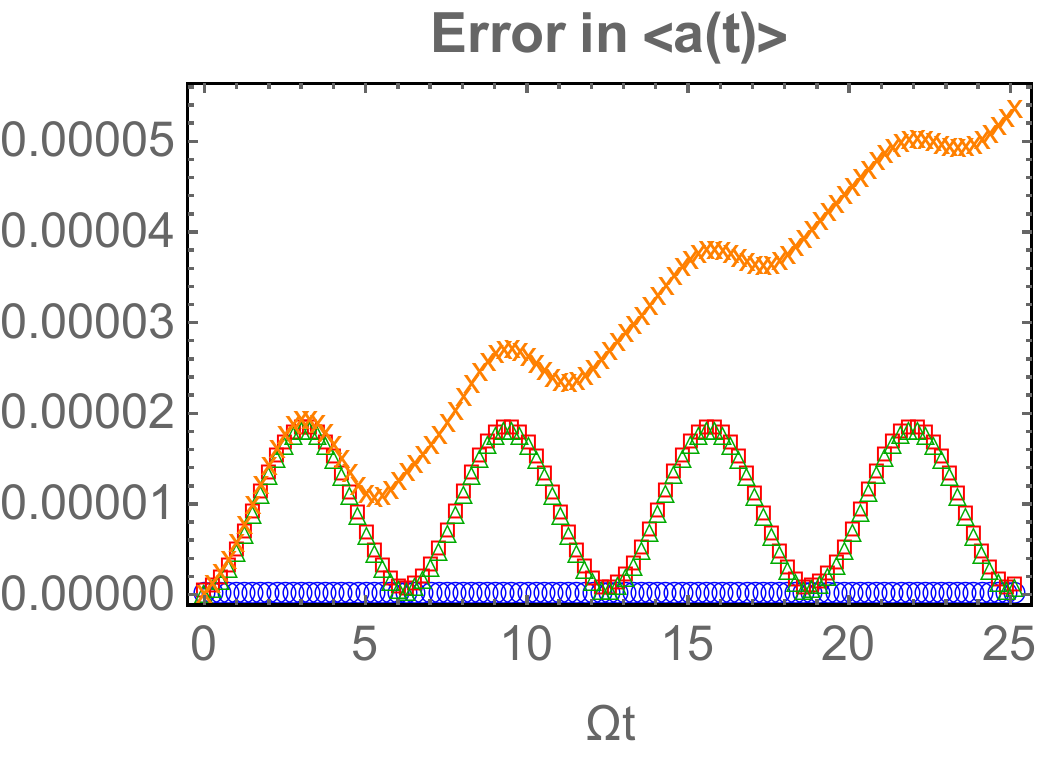}\hspace{.15\linewidth}\includegraphics[height=.25\linewidth,trim=0 0 0 0]{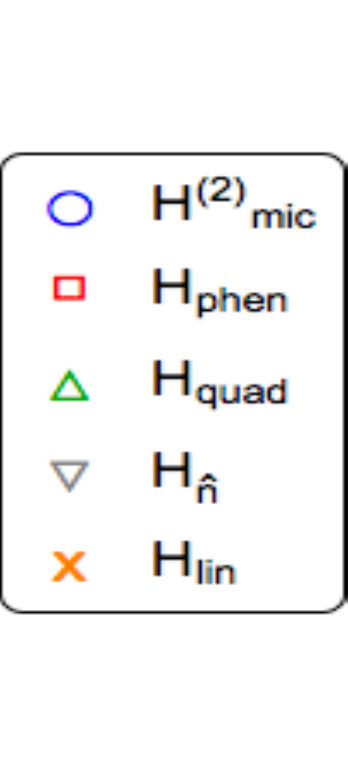}
	\end{flushleft}
	\caption{Errors of the various Hamiltonians in predicting low-energy eigensystems and dynamics for a two-mode optomechanical system. The microscopic model $H_{\sf mic}$ is taken as benchmark (or ``true model"). For these plots we consider a Fabry-Perot optomechanical setup with parameters $\omega_0=10^2\Omega,\, \Omega=10^2g_0.$ The dynamics of $\av{\hat a}$ and $\av{\hat x}$ are calculated for the initial state \eqref{inizialo} with $\alpha=2$. 
		\label{faiga}}
\end{figure*}
%%%%%%%%%
\section{Numerical examples}\label{examples}
{\noindent}So far we have introduced several Hamiltonians that aim to improve the linear model. It is now interesting to compare these refined models in a concrete example: first by examining a few observables that characterize their low-energy eigensystems, then by looking at their predictions for the time evolution of mirror position and field amplitude. 
Since we are looking at small deviations from the predictions of $H_{\sf lin},$ it is practically impossible to compare by eye the plots of absolute quantities. Instead, for any quantity $A$ calculated via a given approximate model, we shall plot the absolute error $|A-A_{\sf mic}|$, where $A_{\sf mic}$ is computed by solving $H_{\sf mic}$ numerically. Note that numerical solutions are adopted also for $H_{\sf phen}$ and $H_{\sf mic}^{(2)}$. For all these cases we wrote a simple Mathematica script to diagonalize the Hamiltonian in a truncated space and hence calculate all quantities of interest. To confirm the solidity of these results we performed the same calculations in Python, which produced visually identical plots. For our purposes truncations of $n_{\sf max}=20$ photons and $m_{\sf max}=30$ phonons were sufficient to obtain convergence, but these numbers rely heavily on the chosen model parameters, and on the number of reliable eigenvalues one is seeking.\\In the displayed examples we set $\Omega=100g_0$, inspired by state-of-the-art experiments featuring $g_0/\Omega\lesssim0.01$ \cite{optorevs1,optorevs2,optorevs3}, while the bare cavity frequency is $\omega_0=100\Omega.$ Qualitatively similar results were obtained by increasing $\omega_0$ to $10^3\Omega$. With our current numerical means we found it difficult to obtain reliable results when increasing $\omega_0$ further: the handling of such cases required prohibitively high numerical precisions. It may be useful to overcome this limitation in the future, in order to address systems where many orders of magnitude separate mechanical and optical frequencies.

\subsection{\bf Low-energy eigensystems}
{\noindent}To compare eigensystems, we focus on the following quantities that depend on the photon number $n$:
\begin{itemize}
\item[(i)] Lowest energy eigenvalue, $E_{n,0}$
\item[(ii)] Average mechanical displacement, $\av{\hat x}_{n,0}$
\item[(iii)] Mechanical uncertainty, $\Delta x_{n,0}$
\end{itemize}
 Note that the eigensystems of Hamiltonians $H_{\sf lin}, H_{\sf phen}^{(2)}, H_{\hat n}$ are \textit{fully characterized} by these quantities. In contrast, $n$ is not a good quantum number for $H_{\sf mic}$ and $H_{\sf mic}^{(2)}$, since these Hamiltonians do not commute with $\hat n$. Yet, in the parameter regimes here considered we were able define an effective photon number $n$ as the integer closest to $\bar n=\av{\hat a^\dagger \hat a}$, since $|n-\bar n|$ remained small in all the reported examples. In turn, this allowed us to identify the lowest eigenvalue at fixed $n$, $E_{n,0}$, as well as the associated eigenstate $\ket{\Psi_{n,0}}$. From these we were able to define, and calculate, quantities (i)-(iii) for the models $H_{\sf mic}$ and $H_{\sf mic}^{(2)}$. The top half of Figure~\ref{faiga} shows that all the approaches considered here provide a significant improvement to the linear model, but at this stage there is no clear winner: while $H_{\sf mic}^{(2)}$ is better at predicting energy eigenvalues, the phenomenological Hamiltonian $H_{\sf phen}$ can predict mechanical equilibrium positions and (most) variances more accurately. In passing, it is interesting to note that $H_{\hat n}$ can be seen as a sort of hybrid model emulating properties of both $H_{\sf quad}$ (equilibrium positions) and $H_{\sf phen}$ (mechanical variances).
\subsection{\bf Dynamics}
{\noindent}To compare dynamical predictions we take the initial state
	\begin{align}
\ket{\Psi(0)}=\ket{\alpha}_a\ket{0}_b, \label{inizialo}
	\end{align}
	where $\ket\alpha_a=e^{-|\alpha|^2/2}\sum_{n}(\alpha^n/n!)\ket{n}_a$ is a coherent state and $\ket{0}_b$ is the bare ground state of the mechanics. It is apparent that the evolution of initial state \eqref{inizialo} is going to involve several photon number sectors even with the number-conserving Hamiltonians. Moreover, according to the linear model, the evolution of this state can induce interesting nonclassical features in the system, such as Wigner-Function negativity and opto-mechanical entanglement \cite{Bose}. Therefore, we think that the initial state \eqref{inizialo} embodies an interesting test-bed for our various approximate Hamiltonians: a reliable description of the subsequent dynamics must capture well the evolution of self-correlations in the cavity field together with entanglement between the two subsystems. In our examples we picked $\alpha=2$, and looked at the accuracy of the various models in predicting the time-dependent averages $\av{\hat x(t)}$ and $\av{\hat a(t)}$: the relevant results are displayed in the bottom half of Figure~\ref{faiga}. From these we infer that $H_{\sf mic}^{(2)}$ has an edge on the other models, as it may be expected from its construction as the `correct' second-order expansion of $H_{\sf mic}$. As a corollary, we argue that the inclusion of counter-rotating terms will eventually become necessary in trying to model optomechanical system with ever higher accuracy.  On the other hand, the good news is that once again the phenomenological approach provides a significant improvement to the linear model (except for a few isolated points in time, where the linear model yields better estimates for $\av{\hat x(t)}$). Furthermore, in the chosen parameter regimes we do not observe significant differences between $H_{\sf phen},$ $H_{\hat n}$ and $H_{\sf quad}$. Under these circumstances, our choice would naturally fall on $H_{\sf quad}$ due to its simple analytical solution. 
\section{Conclusions and outlook}\label{conclusions}
{\noindent}We have proposed and compared several Hamiltonians that go beyond the linear model of optomechanics, featuring corrections that are either phenomenologically or microscopically derived. Using a two-mode truncation of Law's Hamiltonian \cite{CKLaw} as a benchmark, we were able to confirm that the phenomenological approach, common in the literature, is able to provide meaningful corrections to the linear model. A particularly reassuring observation is that, in the regime $\omega\!\gg\!\Omega\!\gg\! g_0$ studied here, these corrections are well captured by the analytically solvable model $H_{\sf quad}$, i.e., a straightforward second-order expansion of the phenomenological Hamiltonian \eqref{Hphenomenal}. However, we also pointed out that the phenomenological approach does not fully capture all second order corrections arising from a microscopic treatment of radiation pressure. In particular we have outlined an important effect, the breakdown of photon number conservation, which can provide non-negligible corrections even in parameter regimes where dynamical-Casimir effects are usually discounted.\\
{\noindent}Our work may benefit future experiments probing the single-photon strong coupling regime or tackling high-precision optomechnical measurements. More importantly, it opens a discussion that may uncover new avenues for the theoretical modelling of optomechanics, beyond the domain of applicability of the linear model. There are several important questions that our work leaves open and that should be addressed in future studies. First, our analysis is confined to a two-mode scenario. In the quest for strong coupling and/or high precision modelling, a multi-mode treatment (including several degrees of freedom for both light and mechanics) may eventually become necessary. Second, we have left system-environment interactions out of our discussion. Due to the presence of three very different energy scales ($\omega_0,\Omega,g_0$), we anticipate that it will be challenging to develop an accurate open-system model for optomechanics that goes beyond the existing phenomenological treatments. Yet, identifying a reliable Hamiltonian is a crucial first step towards the rigorous modelling of an open quantum system \cite{breuer}. Third, we have assumed the bare mechanical potential to be harmonic. Depending on the implementation, anharmonicity in the mirror potential \cite{arqua} may provide new corrections that can be qualitatively different from those studied here. Finally, one may also explore relativistic corrections to the Law Hamiltonian \cite{Relativistic}. The theoretical challenge will ramp up significantly in trying to address any of these limitations of our work, yet it will be a worthwhile pursuit in the quest to model ever more sophisticated optomechanics experiments. 

\section*{Acknowledgments}
{\noindent}We acknowledge fruitful discussions with G. Adesso, F. Armata, A. Armour, S. Bose, L. A. C. Correa, A. Ferraro, M. G. Genoni, O. Hess, M. S. Kim, S. P. Kumar, G. J. Milburn, F. Mintert, A. Nunnenkamp, M. Paternostro, M. Plenio, A. Serafini, H. Ulbricht and especially Daniele Dorigoni. 
T.T. acknowledges financial support from the Foundational Questions Institute (Grant No. FQXi-RFP-1601), and from the University of Nottingham via a Nottingham Research Fellowship. K. S. acknowledges financial support from the University of Nottingham through a Dr Margaret Jackson Summer Bursary Award.
\appendix
\section{Negative energies in the linear model}\label{pathology}
\begin{figure}[t!]
	\includegraphics[width=.9\linewidth]{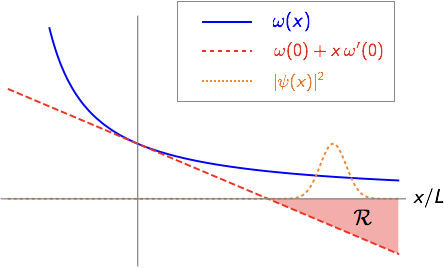}
	\caption{An illustration of the pathology carried by the linear model. While a well-defined optomechanical model features $\omega(x)>0$ for all $x$ (blue continuous curve), the linearization of $\omega(x)$ (red-dashed curve) produces a non-physical region ($\cal R$) where the cavity frequency is negative. It then possible to construct states of arbitrarily low energy by concentrating the mechanical probability density (orange dotted line) in the region $\cal R$, and subsequently populating the cavity with a large number of photons. \label{problem}}
\end{figure}
{\noindent}In this section we shed further light on the pathologies of the linear model of optomechanics, which to the best of our knowledge were first outlined by \cite{patre}. We first take one step back and consider a generic optomechanical Hamiltonian of the form
\begin{align}
H=\omega(\hat x)\left(\hat a^\dagger \hat a+\frac12\right)+\frac{\hat p^2}{2}+V(\hat x),\label{HCK}
\end{align}
where both the frequency dependence of $\omega$ and the mechanical potential $V$ can be left generic at this stage. We will only assume that $V$ is a regular function, so that it cannot encode hard-wall boundary conditions (more on this point below). It is easy to show that, if there is a region for $x$  where $\omega(x)<0$, the Hamiltonian cannot be lower-bounded. Indeed, let us suppose that $\omega(x)$  is not everywhere positive, and that there is a ground state with finite energy $E_0$. For any quantum state it would then follow $E_0\le\av{H}$. In particular let us consider a product state $\ket{n}\ket{\psi}\equiv\ket{n}_a\otimes\ket{\psi}_b$, where $\ket{n}_a$ indicates a fock state of the cavity field, while $\ket{\psi}_b$ is a generic mechanical wavefunction. Taking the expectation value of $H$ we obtain 
\begin{equation}
E_0\le \left(n+\frac12\right) \bra{\psi}{\omega(\hat x)}\ket{\psi}+\bra{\psi}{\left(\frac{\hat p^2}{2}+V(\hat x)\right)}\ket{\psi}.\label{expect_bad}
\end{equation}
If the cavity frequency can assume negative values in a certain region ${\cal R}\equiv\{x\,|\,\omega(x)\!<\!0\}$, it is always possible to obtain $\bra{\psi}{\omega(\hat x)}\ket{\psi}<0$, for example by choosing $\psi(x)\equiv\sprod{x}{\psi}$ as a smooth function with support in $\cal R$ (See Fig.~\ref{problem}). At the same time, the second term on the right-hand side (RHS) of Eq.~\eqref{expect_bad} is a finite constant if the potential does not display singularities. It follows that the expression on the RHS tends to $-\infty$ when $n\to\infty$, so that $E_0$ cannot be finite.

{\noindent}In hindsight this result may appear obvious, as the very concept of a single-mode cavity implicitly assumes $\omega(x)>0$. Yet, the common linear approximation $\omega(x)\simeq\omega_0-G x$ fails to satisfy this requirement. Note that in the above discussion we did not need to specify the functional form of $V,$ hence we can conclude that the outlined pathology cannot be cured by adding anharmonic terms to the mechanical confining potential. The only way to retain a linearised $\omega(x)$, while also ensuring that $H$ in Eq.~\eqref{HCK} is lower bounded, is to impose hard-wall boundary conditions on the oscillator, so that all mechanical wavefunctions are forced to vanish in the region ${\cal R}$.

\end{document}